\documentstyle[eqsecnum,aps,floats]{revtex}
\begin{document}
\tighten
\preprint{SU-GP-97/4-1,hep-th/9705063}
\title{Statistical Effects and the Black Hole/D-brane Correspondence}

\author{Donald Marolf}
\address{201 Physics Building}
\address{Syracuse University,
Syracuse, NY 13244}
\date{\today}
\maketitle

\begin{abstract}

The horizon area and curvature of three-charge BPS black strings
are studied in the D-brane ensemble for the stationary black string.
The charge distributions along the string are used to translate
the classical expressions for the horizon area and curvature of BPS
black strings with waves into operators on the D-brane Hilbert space.
Despite the fact that any `wavy' black string has smaller
horizon area and divergent curvature, the typical values of the
horizon area and effects of the horizon
curvature in the D-brane ensemble deviate negligibly
from those of the original stationary black string
in the limit of large
integer charges.   Whether this holds in general will depend on
certain properties of the quantum bound states.
\end{abstract}



\section{Introduction}
\label{intro}

The correspondence between various classical black hole solutions and
ensembles of D-brane bound states is by now quite well
established \cite{SV,CM,BMPV,MS,JKM,BLMPSV,DMW,DM1,DM2,SM}.
  What is not so clear is the exact meaning of this
correspondence.  For example, controversy continues about the
implications for information loss.  In addition, Myers
\cite{rm} has recently raised the issue of whether the ensemble nature
of the quantum state is essential; that is, whether it might be
impossible to establish a similar correspondence between
a classical black hole and a pure quantum state.  We address this issue below
and, while we will not settle it conclusively,
a framework is provided for its further study.

Let us recall the typical form of results on the black hole/D-brane
correspondence.  For definiteness, we consider the `black holes' of
\cite{MS,LW,CT}
with a 4+1 asymptotically flat space and an internal $T^5$; such objects
correspond to bound states of D-onebranes and D-fivebranes with
internal momentum.  In particular, we will be interested in the case
where one internal $S^1$ is large and the solution may be thought
of as a black {\it string}, with the momentum flowing around this $S^1$.
One first considers the classical black string,
which is defined by a set of `macroscopic parameters' $\{Q_i\}$ -- charges
\cite{SV,CM},
or perhaps long wavelength charge distributions \cite{HM1,HM2}.
Three important such parameters are the total one- and five-brane
charges $(Q_1,Q_5)$ and the momentum quantum number $(N)$.
The black
string solution is chosen to have no structure on smaller scales; that is,
any remaining parameters
$\{q_a\}$ which describe the black string are set
to zero.  Examples of such $\{q_a\}$ are the short wavelength
components of the charge distributions.
One then uses the macroscopic parameters to define an ensemble of
D-brane states.  The macroscopic parameters of such states are required
to agree with the classical black hole,
but their `microscopic structure' is not constrained.
One computes some property of the ensemble such as
the entropy \cite{SV,CM,BMPV,MS,JKM,BLMPSV} or the scattering
cross section \cite{DMW,DM1,DM2,SM}, and compares it with the classical
black hole results, finding agreement in the limit of large charges.

The approach below is somewhat different.
Our goal is to identify a quantum operator $A$ whose
classical limit yields the horizon area of the black hole and
a family of operators $R_{\alpha \beta \gamma \delta} (x)$ whose
classical limits give the curvature tensor in some coordinate system.
These will involve both the macroscopic $\{Q_i\}$ and the
microscopic $\{q_a\}$.
We then use these operators to study statistical mechanical
effects on the horizon area and curvatures.  A quantum
treatment is necessary since the bound states form a field theory
with the corresponding divergences in classical statistical mechanics.

We begin by taking seriously
the conjecture that one can establish a correspondence between
(suitable limits of) individual D-brane states and classical
black hole solutions.  This is consistent with the usual
picture of a classical state corresponding to a
limit of quantum states, as $\hbar \rightarrow 0$.
In this case, any function on the space of classical solutions should
arise as the classical limit of an operator on the quantum state
space.    We will use the classical expressions for $A$ and
$R_{\alpha \beta \gamma \delta} (x)$ in terms of the charge
distributions to motivate a choice of quantum operator.
The expected values of $A$ and $R$ will then be computed in the D-brane
ensemble and compared with the stationary classical black string.
Note that we do not require the weakly coupled D-branes to have
a horizon of area $A$ or curvature $R_{\alpha \beta \gamma \delta}(x)$
in any physical sense;  the point is simply that, if the D-brane
ensemble successfully characterizes the charge distribution fluctuations
of the black string, such operators will provide the  effects of these
fluctuations on the area and curvatures.
We consider the case where the `macroscopic parameters' $\{Q_i\}$
are just the integer charges while the microscopic parameters
$\{q_a\}$ include all information about the distribution of this charge.

One typically expects that any unconstrained quantity (such as $q_a$)
will average to zero and that fluctuations are small in the thermodynamic
limit.  As a result, the expected values of $A$ and $R$ should
not differ significantly from their values for the original
black string.  However,  the
area operator has the property that it is equally  sensitive to
fluctuations on all distance scales.  In fact,
the expected value of $A$ will receive a correction from each
$q_a$ that is nonzero and these corrections are additive.
The curvature depends even more
dramatically on many of the microscopic parameters $q_a$.  As shown in
\cite{HY}, any inhomogeneities in the distribution of longitudinal
momentum along
the string results in a singular curvature at the horizon.  Thus, this
curvature in general diverges when $q_a \neq 0$.  However, the
singularity is weak and its physical effects are finite.  The point of our
study is to verify that the fluctuations are small enough to have
negligible effects even on these sensitive characteristics.  We will
find that the resulting deviations are suppressed by powers
of $Q_1$ and $Q_5$, but not by powers of $N$.  As a result, the
horizon area and curvatures of a stationary black string correspond
well to $\langle A \rangle$ and $\langle R \rangle$ not just for the
D-brane ensemble as a whole, but also for a typical state.

In the course of our argument, a number of subtleties will arise
in relating the classical black strings to quantum states.
They are connected to the question of how to choose the parameters
$q_a$ and the function $A(Q_i,q_a)$.  A working hypothesis is
stated, and arguments are given in its favor.  A complete
understanding of this issue is likely to elucidate many features of
the black hole/D-brane correspondence.

The structure of this paper is as follows.  Section \ref{class} reviews
certain properties of the classical supergravity solutions corresponding to
six dimensional BPS black strings, providing a slight generalization of
previous work.  These results will be used to define
the D-brane `area
operator' and `curvature operator'
as it is in terms of such asymptotic properties that the
black hole/D-brane correspondence seems most direct.  Section \ref{quant}
studies the area operator while
section \ref{sings} studies the
curvatures.   We close with  a discussion of results and certain
subtleties in section \ref{concl}.

\section{The Space of Classical Solutions}

\label{class}

Recall that our goal is to use the classical expression for the horizon area
and curvature in terms of the charge densities to define
quantum operators on the D-brane Hilbert space.
As a result, we should
consider a space of classical solutions and a set of D-brane states
such that each classical solution is associated with a suitable
limit of our D-brane states (or ensembles of such states) and
such that each D-brane state takes part in this limit.

We focus on the class of solutions known as six-dimensional BPS
black strings with traveling waves \cite{LW,CT,AT}.  Such solutions
generalize the more standard stationary and translationally invariant solution
\cite{HMS,CY} and have been shown to correspond to ensembles of D-brane
states \cite{HM1,HM2}.  In particular, the associated D-brane
ensembles are {\it subsets} of the ensemble usually used to describe
the stationary black string \cite{MS}.  As a result, despite the singularities
found in \cite{HY}, such solutions must be included in our discussion.

Ideally, we would like to consider the space of {\it all} BPS
solutions.  Then, assuming that A) the resulting horizon area and curvatures
can be written as a function of the asymptotic charge
distributions and B) a correspondence between the asymptotic
charge distributions of the classical BPS solutions and the D-brane
states can be made, we would then use these charges to construct operators
on the D-brane Hilbert space.
However, to even state precisely just
what is meant by `all BPS solutions' would require a
detailed understanding of the classical limit of all
BPS string states.
As a result, we shall take the much more modest approach
of using a set of classical solutions which has already been studied in
the literature to refine the area only
a single step beyond the most naive $A= (4G_N) 2 \pi \sqrt{Q_1Q_5N}$,
and similarly for the curvatures.

Specifically, we consider the BPS black strings
of \cite{HM1,HM2}, generalized to allow spatially varying moduli
and the simultaneous presence of waves carrying angular momentum,
those carrying
momentum around the $T^4$, and the usual `longitudinal waves.'
Recall that the low energy action for type
IIB string theory (in the Einstein frame) contains the terms

\begin{equation}
S = {1 \over {16 \pi G_{10}}} \int dx^{10} \sqrt{-g} \left(
R - {1 \over 2}(\nabla \phi)^2 - {1 \over {12}} e^\phi {\cal H}^2 \right)
\end{equation}
where $\phi$ is the dilaton and ${\cal H}$ is the Ramond-Ramond three form.
We are interested in solutions for which there
are $4+1$ asymptotically
flat dimensions ($x^a,t$), one `large' internal  $S^1$ ($z$) of length $L$
(at infinity),
 and four `small' internal
dimensions forming a $T^4$ ($y^i$) of volume ${\sf V} = (2\pi)^4 V$ (at
infinity).
There is a D-onebrane charge in the
large $S^1$ direction and a D-fivebrane charge around the full internal
$T^5$.  The moduli satisfy $L,V^{1/4} \gg l_{string}$ so that the
classical supergravity description is appropriate.
We require the horizon to be `hole-shaped' ($S^3 \times
{\bf R}$) as viewed
from the asymptotically flat space, and to have topology $S^3 \times T^5 \times
{\bf R}$
in the full ten dimensional solution.   We consider solutions
in which the only complications added to the simplest black string
solutions are certain `waves' which run around the large $S^1$.
In particular, we take our solutions (in the Einstein frame) to be of the form

\begin{equation}
\label{metric}
ds^2 = H_1^{1/4} H_5^{3/4} \left[ {{du} \over {H_1 H_5}}
(-dv  + K du + 2 A_i dy^i + 2A_a dx^a)
+ {{dy^i dy_i} \over {H_5}} + dx^adx_a \right]
\end{equation}
\begin{equation}
\label{dil}
e^{-2 \phi} = {{H_5} \over {H_1}}
\end{equation}
\begin{equation}
\label{gf}
{\cal H}_{auv} = H_1^{-2} \partial_a H_1, \ \ {\cal H}_{aub} =
2\partial_{[a}A_{b]}, \ \ {\cal H}_{iuj} = 2 \partial_{[i}A_{j]}, \ \
{\cal H}_{abc} = -
\epsilon_{abcd} \partial^d H_5
\end{equation}
where $H_1 = 1 + {{r_1^2} \over {r^2}}$, $H_5 = 1 + {{r_5^2} \over {r^2}}$,
$dx^adx_a = dr^2 + r^2 (d\theta^2 + \sin^2 \theta d\varphi^2
+ \cos^2 \theta d\psi^2)$ and
 $K$, $A_a$, and $A_i$ are functions of $x,y,u$ but not
$v$.  The indices $i,a$ are raised and lowered with
the Euclidean metrics $\delta_{ij}$ and $\delta_{ab}$ and both run over
$\{1,2,3,4\}$.
Here the physical values
$r_1,r_5$ of the one- and fivebrane charges as related to the quantized
integer charges $Q_1$ and $Q_5$ through
\begin{eqnarray}
Q_1 &=& {{V r_1^2} \over g} \cr
Q_5 &=& {{r_5^2 } \over g}
\end{eqnarray}
where $g$ is the string coupling constant.  The horizon lies at $r=0$.

Note that the asymptotic
values of $K$, $A_i$, and $A_a$ determine the total momenta carried
by the black string.  Due to the null translational symmetry,
it seems natural to interpret $K$, $A_i$, and $A_a$ as defining momentum
{\it densities}
in the $z$ and $x^i$ directions
as functions of $u$, an angular momentum density, and even
higher multipole moments
of the momentum and angular momentum densities\footnote{See, for example,
\cite{mult} for a construction of certain multipole moments in stationary
spacetimes.} on the $S^3$ and $T^4$.
We take
\begin{eqnarray}
\label{fields}
K &=& {{p(u)} \over {r^2}} \cr
A_i &=& - {{p_i(u)} \over {r^2}} \cr
A_a dx^a &=& {{\gamma (u)} \over {r^2}} (\sin^2 \theta d \varphi -
\cos^2 \theta d \psi)
\end{eqnarray}
so that the momentum and angular momentum densities are

\begin{equation}
{{dP_z} \over {du}} =  \kappa^{-2}  p
\end{equation}
\begin{equation}
{{dP_i} \over {du}} =  \kappa^{-2}  p_i
\end{equation}
\begin{equation}
{{dJ_\varphi} \over {du}} = - {{dJ_\psi} \over {du}} = \kappa^{-2}
\gamma(u)
\end{equation}
\begin{equation}
{{dE} \over {du}} =  \kappa^{-2} ( r_1^2 + r_5^2 + p)
\end{equation}
where we have introduced
\begin{equation}
\kappa^2 = {{4 G_{10}} \over {\pi {\sf V}}} = {{2 \pi g^2} \over V}.
\end{equation}
This is essentially the same as the case studied in \cite{HM1,HM2},
though our parametrization of the waves is slightly different.
Note that the momentum {\it densities} are conserved on this
class of solutions since ${{\partial} \over {\partial v}}$ is
a (null) Killing field.

The reader may notice that we have {\it not} allowed our black string
to carry what were called `external waves' in \cite{HM1,HM2}.  These
are the waves which give the string a nonzero density of momentum in
the $x^a$ directions and which can be thought of as transverse
oscillations of the black string.  Such waves have been excluded
since they involve oscillations of both the one-brane and five-brane
degrees of freedom (the black string oscillates as a whole).  Given
that a complete description of the five-brane degrees of freedom is
beyond the scope of this work, leaving out the external waves is a natural
choice.

Since we have allowed only low order multipole terms in (\ref{fields}),
a slight extension of \cite{HM1,HM2} shows that
the metric is $C^0$ at
the horizon so that the
horizon area is
well defined.  It takes the form
\begin{equation}
\label{sigma}
A = 2 \pi^2 r_1^2r_5^2 {\sf V} \int \ du \ \sigma(u)
\end{equation}
where $\sigma(u)$ is a periodic solution of
\begin{equation}
r_1^{-2} r_5^{-2} [p(u) - p_i(u) p^i(u)/r_1^2 - \gamma^2(u)/r_1^2r_5^2]
 = {{\partial} \over {\partial u}}{\sigma} + \sigma^2.
\end{equation}
Thus, for this class of solutions, we have expressed
the horizon area in terms of the asymptotic charge distribution.
For `slowly varying waves' satisfying
\begin{equation}
\label{slow}
{\partial \over {\partial u}} [ p - p_ip^i/r_1^2 -\gamma^2(u)/r_1^2r_5^2
 ] \ll [p - p_ip^i/r_1^2 - \gamma^2/r_1^2r_5^2]^{3/2},
\end{equation}
a good approximation is
\begin{equation}
\label{approx}
\sigma
= {1 \over {r_1r_5}} \sqrt{p(u) - p_i(u)p^i(u)/r_1^2 -
\gamma^2(u)/r_1^2r_5^2}.
\end{equation}  In the case of a small (and slowly varying)
deviation from the stationary
solution $p(u)=p$, $p_i(u) = 0$, $\gamma = 0$, the average
value $\overline{\sigma}$ of $\sigma$ reduces to
\begin{eqnarray}
\label{expand}
\overline \sigma
&=& {1 \over {r_1r_5}} \sqrt{p} \Bigl (1 +  L^{-1} p^{-2}
\int \ du \Bigl[ p \  \delta p(u) -
{1\over 4} (\delta p(u))^2 - {1 \over 2} p_i(u) p^i(u)/r_1^2  -
{1 \over 2}
\gamma^2(u)/r^2_1r^2_5 \Bigr] \cr &+& O(\delta p^3) + O(p_i^4) + O(\gamma^4)
\Bigr),
\end{eqnarray}
where $\delta p (u) = p(u) - p$.  This is  the expression which will
be used in section \ref{quant}.
It is natural to choose $p$ such that $\int du \ \delta p = 0$,
so that the lowest order deviation is the quadratic term.
Our study of the area operator reduces to computing
the fluctuations $\delta p(u)$, $p_i(u)$, $\gamma(u)$ of the
charge distributions for our black string.

At this point, the question should be asked if the space of solutions
given by \ref{metric}, \ref{dil}, \ref{gf}, \ref{fields} is
in fact large enough, or must be further enlarged before a meaningful
comparison with the D-brane Hilbert space can be made.
A conclusive statement is beyond the scope of this work, but let
us at least address the most obvious possibility of considering
higher multipole terms in $K$, $A_i$, and $A_a$.  We
recall first that, in order for (\ref{metric}), (\ref{dil}), and (\ref{gf})
to be a solution, the
fields $K$, $A_a$, and $A_i$ are typically constrained to satisfy
some elliptic set of equations.  For example, when $A_i$ is divergence
free and $A_a=0$, we must have \cite{HM3}

\begin{eqnarray}
(\partial_x^2 + H_5 \partial_y^2)K &=& 0 \cr
(\partial_x^2 + H_5 \partial_y^2)A_i &=& 0
\end{eqnarray}
where $\partial_x^2$ and $\partial_y^2$ denote the usual flat-space
Laplacians in $x$ and $y$.  For the purposes of this discussion, 
let us impose the boundary conditions that $K$ and $A_i$ are
bounded near infinity so that we do not modify the asymptotic
structure of the spacetime.  While the general form of these equations
has not yet been derived, one expects that the rough picture developed
in \cite{HM3,KMR} will continue to hold with the fields $K$, $A_i$, and
$A_a$ being entirely determined by the various multipole moments
of the charge distributions and having the property
that, whenever any of the dipole or higher moments is nonzero, the
solution is singular on the horizon.  This was shown in
\cite{HM1,KMR} to be precisely true of the field $K$ and a short calculation
based on the result of \cite{KMR} shows that, at least for the
case studied there, this singularity is `strong' in the sense that both
the once and twice integrated curvatures also diverge at the horizon.
As a result, the singularity produces an infinite distortion of any
object attempting to cross the horizon\footnote{Recall that this was not
the case for the singularity discovered in \cite{HY} associated with
monopole waves.}.

It seems reasonable to exclude strongly singular solutions, although a
fully convincing argument would have to follow from some careful
analysis.  In an earlier version of this paper it was stated that such
an analysis would have to center on the details of the BPS D-brane
states.  However, new evidence \cite{SSDM} suggests that a study of
classical solutions may be sufficient.  Due to the `deep throat' of the
extremal black string, an inhomogeneity in the
distribution of one-brane charge causes a much smaller effect on the asymptotic
fields than one would naively expect when the charge is placed
close to the horizon.  The result is that, despite the fact that the
$T^4$'s near the horizon have finite size, an inhomogeneity of finite
`intrinsic' strength located at the horizon produces {\it zero} effect
on the external spacetime.  We suspect that the same is true for
other types of charge.  Details and further results will appear in
\cite{SSDM}.

It is important to note, however,
that if such higher multipole solutions {\it are} relevant to the
D-brane ensemble,
then due to their strong singularities they should
dominate any discussion of the effects on an
object passing through the horizon.  Moreover, because the $du^2$ term
diverges, they appear
to have infinite horizon area.
Thus, if such solutions are included, the expected
area and curvatures in the ensemble should be vastly different
than the area and curvatures of the original classical solution. We
adopt the working hypothesis that all BPS D-brane states in which
the fivebranes do not oscillate correspond to
`monopole' solutions and exclude other solutions from our study.

As a final comment, recall that we have already excluded the solutions
with `external' waves as they involve oscillations of the five-brane
in the D-brane description.  Such waves are slightly different
than the ones considered here as they are associated with a charge
(momentum in the $x^a$ direction) which is a vector from the point of
view of the noncompact spacetime.  In contrast, the waves considered
here correspond to scalar charges in this sense.  Note that, if we
take as our basic principle that the allowed fields are those that
correspond to a constant charge density over the $S^3 \times T^4$ in the 
flat space limit, the corresponding external waves are {\it not}
$SO(4)$ invariant.  Thus, $l=1$ external waves are the analogue of the
monopole waves considered here, while it is only the $l \ge 2$ modes
that correspond to `higher multipole waves.'  Since $l=1$
external waves are not strongly singular, this is consistent
with our overall point of view.

\section{Area deviations and fluctuations}
\label{quant}

We now discuss the quantum D-brane states and write down an operator
$A$ on the D-brane Hilbert space which corresponds to the black hole
horizon area.  The curvatures will be studied in section \ref{sings}.
We now restrict to the regime $ \sqrt{p} \ll r_1,r_5$ so that we may use the
the effective string
model of \cite{JL,HW}.  Thus, the ensemble for the homogeneous
black string contains those states
on a single string of length $\tilde{L} = Q_1Q_5L$ and
tension $T = 1/2 \pi g Q_5$ having total momentum $P$.
Four right moving bosonic fields and four right moving fermionic fields live
on this string.  However, as our goal is only to estimate the order of
magnitude of certain effects
and to show that they are suppressed by positive powers of $Q_1$ and $Q_5$,
it will not be necessary to keep track of all the fields.
Let us consider only a single  right moving
bosonic field $\chi(u)$; the behavior of the fermion fields is much the same.
We may think of the field $\chi$ as representing the displacement of the
effective string in one of the four internal directions, say $y^1$.

As a result, our system is described by the action

\begin{equation}
S  =  - {1 \over {4 \pi g Q_5}} \int (\partial \chi)^2
\end{equation}
and has a momentum density
\begin{equation}
T_{++} = (\partial_+ \chi)^2 /(2 \pi g Q_5)
\end{equation}
along the string and a momentum density $\partial_+ \chi/(2 \pi g Q_5)$
in the direction of the transverse oscillations (the $y^1$ direction).
Here we use coordinates $\sigma_-, \sigma_+$ along the string worldsheet.
As usual, we take the momentum density to be normal ordered\footnote{Due
to supersymmetry, this would be unnecessary if the fermions were
explicitly included.}.
We will use the mode expansion
\begin{equation}
\partial_+ \chi = {{\pi \sqrt{2g Q_5}}
\over {\tilde{L}}} \sum_{n=-\infty}^{\infty}
\alpha_n e^{-2\pi i n \sigma_+ /\tilde{L}}
\end{equation}
with $[\alpha_m, \alpha_n] = m \delta_{-m,n}$.

Recalling that the effective string wraps $Q_1Q_5$ times around the
compact direction, values of the worldsheet coordinate $\sigma_+$
which differ by integer multiples of $L$
correspond to the same value of the spacetime coordinate $u$.
As a result,  parameters in the black string solution may then be identified
with quantum fields on the effective string though
\begin{equation}
p(u) = {{\kappa^2 \pi} \over {\tilde{L}^2}}\sum_{k=1}^{Q_1Q_5}
\sum_{n=-\infty}^{\infty}
\sum_{m=-\infty}^{\infty} e^{-2\pi i(n+m) (u+kL)/\tilde{L}} :\alpha_m \alpha_n:
\end{equation}
and
\begin{equation}
p_i = {{\kappa^2} \over {\sqrt{2 g Q_5}}}
\sum_{k+1}^{Q_1Q_5} \sum_{n=-\infty}^{\infty}
\alpha_n e^{-2\pi i n (u+kL) /\tilde{L}}
\end{equation}
where $::$ denotes the normal ordering.  

Recall that our plan is
to substitute this expression into equation (2.15), and thereby to
define a quantum area operator.  To do so, we will have to consider
products of the $p(u)$ with itself at the same point.  We choose
to deal with such products by normal ordering them wherever they occur.
A priori, it is not clear that this is the correct approach.  However, 
as our goal is to compute statistical effects (and not quantum effects), 
we feel that such a treatment is normal ordering is sufficient for this purpose.

To show that only small corrections result from taking the $u$-dependence
into account, we need only study quadratic combinations of these
operators in the appropriate ensemble.
The ensemble to be used is the usual one placed in correspondence
with the stationary black hole:  the microcanonical ensemble with
total momentum $P = P_z$.  Rather than explicitly compute $ \langle :
\delta p^2 : \rangle$,
$\langle :p_ip^i: \rangle $, $\langle: \gamma^2 :\rangle$
 in the entire ensemble, we will make use of the
equipartition theorem to simplify the analysis.   A short derivation
of the equipartition theorem in this context proceeds as follows.

Recall that the entropy of the microcanonical ensemble is of order
$S(P) \sim \sqrt{Q_1Q_5N}$ where $N = PL/\hbar$. Let $P'$ be the momentum
of an arbitrary state in the ensemble.  For large
$N$, the ensemble with {\it fixed} momentum $P' = P$ is equivalent to
an ensemble which allows any state with momentum $P' \le P$;
such an ensemble is overwhelmingly dominated by the states with maximal
momentum.  Similarly, for large $N$, such an ensemble is equivalent to
one in which states are included with a probabilistic weight
proportional to
$e^{-\beta P'}$, where $\beta$ is chosen by fixing the peak of
$e^{-\beta P' + S(P')}$ to be at $P'=P$.  Since the total momentum $P'$
of the state is just the sum of the momenta carried by the individual
modes, the weight separates into a product of weights $e^{-\beta P_n}$
for each mode $n$, where $P_n = {{2 \pi n } \over {\tilde{L}}} N_n$ and
$N_n$ is the occupation number of the $n$th mode.  For any mode $n$ with
small energy and large occupation number, the classical limit applies and
the momentum carried by
that mode is $P_n = \beta^{-1}$, independent of $n$ as
predicted by the equipartition theorem.  Higher
modes receive corrections so that they carry momentum $P_n =
\beta^{-1} (1 + O(1))$ and, above some cutoff $n_{max}$, the momentum
is essentially zero.  Thus, a sufficient approximation for our
purposes is to set $P_n = \beta^{-1} (1 + O(1))$ for $n \le n_{max}$ and
$P_n = 0$ for $n > n_{max}$.
Setting $\tilde{N} = Q_1Q_5 N$ and taking
$nN_n = n_{max}N_{n_{max}}$ for $n \le n_{max}$, we
must have $n_{max} P_n \sim n^2_{max} N_{n_{max}}
\sim \tilde{N}$. Since
the cutoff must occur at some  $N_{n_{max}} = \gamma^2$ of order $1$,
we have $n_{max}
\sim \sqrt{\tilde N}/\gamma \sim \sqrt{\tilde N}$.   It
is perhaps reassuring to note that this result can then be used to calculate
the entropy of the ensemble through $dS =  \beta dP =
{1 \over {\gamma \sqrt{\tilde N}}} d \tilde N$
yielding $S =  \gamma^{-1} \sqrt{\tilde{N}}$.  If we wish, we may then
use this to maximize the validity of our approximation by choosing
$\gamma = 6/(2 \pi)^2$.

  This result will allow us to easily
estimate the difference between the actual expectation value of the
area in the ensemble and the area of the stationary black string.
Let us first consider the term coming from
deviations in the longitudinal
momentum.  Expanding the momentum as
\begin{equation}
p(u) = {1 \over L} \sum_{n = -\infty}^{\infty}  p_n e^{2\pi i n u /L}
\end{equation}
and comparing against
the stationary black string with $p = L^{-1} \int \ du \ p(u)$,
we need only show that the quantity
\begin{equation}
\sum_{k \neq 0}  {{<:p_k p_{-k}:>} \over {<:p_0^2:>}}
\end{equation}
is small, where $p_0 = {{2 \pi \kappa^2 \tilde N} \over {\tilde L}}$.
In a state with occupation numbers $\{N_n\}$,
the expectation value of $p_kp_{-k}$ is
\begin{eqnarray}
\langle \{N_n \} | :p_k p_{-k}: | \{ N_n \} \rangle
&=& {{8 \kappa^4 \pi^2 } \over
{{\tilde L}^2}} \Bigl( \sum_{Q_1Q_5k/2 \ge n \ge 0}
(n N_n) (Q_1Q_5k-n) (N_{Q_1Q_5k-n}) g_k(n) \cr &+&
 \sum_{n < 0} (-n) N_{-n}
(Q_1Q_5k-n) N_{k-n} \Bigr)
\end{eqnarray}
where $g_k(n) = 1/8$ for $n=Q_1Q_5k/2$ and $g_k(n) =1$ otherwise.  Using
our equipartition results that $nN_n \sim \sqrt{\tilde N}$ for $n \le
n_{max} \sim \sqrt{\tilde N}$, this gives
\begin{equation}
{{\langle  :p_k p_{-k}:  \rangle} \over {\langle :p_0^2: \rangle}}
 \sim  {\tilde N}^{-1/2}
\end{equation}
for $Q_1Q_5k \sim n_{max}$ while $\langle :p_k p_{-k}:
\rangle \sim 0$ for
$k \gg n_{max}/Q_1Q_5$.  As a result, fluctuations in the longitudinal momentum
cause a deviation
from the area of the stationary black hole by a fractional amount:
\begin{equation}
\sum_k {{\langle :p_k p_{-k}: \rangle } \over {\langle :p_0^2: \rangle }}
\sim {1 \over {Q_1 Q_5}}.
\end{equation}

The contributions from the $:p_ip^i:$ and $:\gamma^2:$ terms are even smaller.
Although the angular momentum is carried entirely by the fermions,
the calculation of $\langle \gamma^2 \rangle$ remains much the same.
Instead of the equipartition theorem, the Fermi sea approximation
of $N_n =1$ for $n \le n_{max}$, $N_n = 0$ for $n > n_{max}$ is useful
in this context.  The results are
\begin{equation}
{{\langle :p_i p^i: \rangle } \over {p r_1^2} } \sim {{Q_1} \over 
{(Q_1 Q_5)^2}};
\ \ {{\langle :\gamma^2: \rangle}
\over {p r_1^2 r_5^2}} \sim {1 \over {Q_1Q_5}}.
\end{equation}

This sort of argument can also be used to justify our use of the slowly
varying approximation (\ref{slow}) for the area.
Another calculation along the lines of those above
yields
\begin{equation}
\langle :\dot{p}^2: \rangle \sim  {{(\kappa^2 \pi )^2 N^3}
\over {L^6 Q_1^2 Q_5^2}}
\end{equation}
while
\begin{equation}
{{\langle :\dot{p}_i \dot{p}^i: \rangle} \over {r_1^2}} \sim {{\kappa^2 N^2 }
\over {L^4 (Q_1 Q_5)^2 Q_5}};
 \ \  {{\langle :\dot{\gamma}^2: \rangle} \over {r_1^2 r_5^2}}
\sim {{\kappa^2 N^2} \over {L^4 (Q_1Q_5)^2}}
\end{equation}
so that $\dot{p}_{rms} \gg {{ \langle: \gamma \dot{\gamma}: \rangle } \over
{r_1^2 r_5^2}} \gg {{\langle :p_i \dot{p}^i: \rangle } \over {r_1^2}}$ and
\begin{equation}
{{\dot{\sigma}_{rms}} \over {\sigma^2} } \sim {1 \over {\sqrt{Q_1Q_5}}} \ll 1.
\end{equation}

We shall not explicitly compute the fluctuations in the area operator, but
we note that our computations of $\langle \delta p^2 \rangle$,
$\langle p_ip^i\rangle$, and $\langle \gamma^2 \rangle$  are
already measures of the size of fluctuations in our ensemble, and
we have $\delta \sigma_{rms} \sim \langle \sigma \rangle /\sqrt{Q_1Q_5}$.
In this sense then, the charge distributions have only small fluctuations.
As a result, a typical state has momentum density $p + O(\delta p_{rms})$
and an area extremely close to that of the stationary black string.

\section{The size of the singularities}
\label{sings}

The discussion so far has focused on the `area operator' that one
might define by comparison with the classical solutions (\ref{metric}).
This quantity was of interest both because of its
prominence in the discussion of black hole entropy and because it
was sensitive to short wavelength fluctuations.
There is, however, another characteristic of the black
string solutions which displays an even more startling sensitivity to
inhomogeneities -- this is the curvature at the black string horizon.
As shown in \cite{HY}, the curvature components in a parallel propagated
orthonormal frame diverge at the horizon whenever the BPS black string is
not exactly translationally invariant. Thus, if one were to define
a `horizon curvature operator' in the D-brane Hilbert space by
looking at the dependence of the horizon curvature on the asymptotic
momentum distributions $p(u)$, $p_i(u)$, $\gamma(u)$, it would diverge on all
states in the microcanonical ensemble, in marked contrast to the
finite curvature of the stationary black hole solution.  Does it follow
that any object approaching a `real' macroscopic black string would
be ripped apart by the large curvatures present in its microscopic
quantum description?

In fact, it was already shown in \cite{HY} that the singularity
at the horizon is of a relatively mild sort.  Although the curvature
itself diverges, the total distortion of, say, a set of uncharged
test
particles passing through the horizon is given by the twice integrated
curvature which is finite.  We shall also see below that the
relative velocities induced between two such test particles remain bounded
as the horizon is approached.  Thus, it is natural to ask the following
questions:  1) For freely falling test particles, how
do the relative velocities that two dust particles obtain from the
inhomogeneous black string compare with those imparted by the stationary
black string?  How does the
total tidal distortion caused by the inhomogeneities compare with the
distortion that would be induced by a stationary black string?  2)
We may also wish to consider test particles which
are not freely falling, but instead remain at a constant value of the radial
coordinate outside the horizon\footnote{Thanks
to Rob Myers for raising this issue.}.
This is the worldline followed by an extremal test particle
with the same sign of the charge as the black hole.
We should therefore ask, ``how close does
such a test particle have to be to the horizon before the divergent
tidal force from the wave becomes comparable to the tidal
forces present in the stationary black hole?''  Again using the asymptotic
charges to map the appropriate functions on the space of classical solutions
to operators on our D-brane Hilbert space, we will find that,
for the microcanonical ensemble which corresponds to the stationary black hole,
the effects from the waves are smaller than the
stationary effects by powers of $1/{Q_1 Q_5}$.  Since we
use the results of \cite{HY}, for the rest of this section we
shall restrict ourselves to the case $r_1 = r_5 = r_0$,
$\gamma(u)=0$, $p_i(u) = 0$, considered there.  While we must
restrict to the case $\sqrt{p} \ll r_1,r_5$ to use the fluctuation
results from section \ref{quant}, this approximation has not
been explicitly used below; all classical terms of any order in
$\sqrt{p}/r_0$  have been kept.

\subsection {Comparing Curvatures}

Let us begin by answering the second question -- just how close to
the horizon must we be
for the curvature caused by waves
to become comparable to the
curvature of the stationary black hole?  When written in
terms of the one-form basis used in \cite{HY}, the divergent
curvature terms have the form

\begin{equation}
R  \sim {{\dot{\sigma} r_0^2 } \over {r^2}} \sim {{\dot p} \over
{r^2 {\sqrt p}}}
\end{equation}
where we have restored the dimensional factors and translated the
expression in \cite{HY} into the coordinates used here.  We
must, of course, know something about the curvatures produced by
a stationary black hole as well.  It turns out \cite{PHY}
that, for the stationary
black hole, the leading order term
near the horizon is
$p/r_0^4$.  Again using the
root mean square value of $\dot{p}$, the term from the inhomogeneities
dominates the stationary curvature only when
\begin{equation}
r  \sim \left( {{\dot{p}_{rms}}  \over {p^{3/2}}} r_0^4 \right)^{1/2}
\sim r_0 (Q_1 Q_5)^{-1/4}.
\end{equation}
Let us denote this value of $r$ by $r_A$.
While we have derived this result using the curvature components in
a particular basis, we may expect a similar result in general as
both the wave-induced and stationary parts
of the curvature are associated with curvature components proportional
to the same null one-form $du$.

It is interesting to note that, when written in terms of the charges
and moduli, $r_0 = (Q_1 Q_5)^{1/4} \sqrt g/V^{1/4}$.  Thus, the
transition occurs at
\begin{equation}
r \sim r_A = \sqrt{g} V^{-1/4}
\end{equation}
where factors of the string length have been set to one.
For the classical supergravity description to be valid, we should have
small coupling and large $T^4$ so that $r_A \ll l_{string}$.
Although $r=r_A$ is still an infinite proper distance from the horizon,
the placement of a particle at $r=r_A$
requires extreme care.
For example, the corresponding redshift from infinity is
$r_A/r_0$, so an uncharged particle dropped in from infinity
would have to shed $(1-r_A/r_0)$ of its mass to instantaneously come
to rest at $r=r_A$.
Also, while it is not clear to what extent quantum effects become
relevant outside the horizon of an extremal black hole\footnote{
In particular, it is interesting to note that, for an
extremal black {\it hole}, the proper acceleration required for a test particle
to travel along an orbit of the Killing field ${\partial \over {\partial t}}$
remains bounded and approaches an asymptotic value of order $1/r_0$ as
$r \rightarrow 0$.  This observation was also made in \cite{PC}.
Similarly, a geodesic which is initially tangent to
${{\partial} \over {\partial t}}$ requires a proper time or order $r_0$
to fall across the horizon, even if it begins its journey at very
small $r$.
Due to its ergosphere, a black string with momentum is more complicated, but it
remains true that a test particle requires only an acceleration of order
$1/r_0$ to remain at any
 value of $r$.},
if they are relevant anywhere, they are likely
to be relevant at such a value of
$r$.

\subsection{Curvature Effects}

Let us now consider a set of test particles whose worldlines are not
so finely tuned, but which fall across the horizon on geodesics.
We would like to compute the relative velocity and total distortion
induced by the inhomogeneities.
Since the
stationary curvature terms dominate the terms from the time
dependence
outside $r = r_A$, any distortion produced by the
inhomogeneities before reaching this value of $r$ will be negligible compared
to that induced by the stationary terms.  As a result, we need only
follow our test particles from $r_A$ to the horizon.

The relative velocity $v$ induced between two test particles
separated by a small distance $l$ is given by integrating the curvature once
along the worldline.  This is most easily accomplished by using coordinates
$(U,V,q,\theta)$ based on those of \cite{HM1,HY}; details of the
coordinate system are given in appendix A.  What is important
about this coordinate system is that the horizon is just the surface
$U=0$ and that the metric is $C^0$ there.  The coordinates
$\theta$ on the three-sphere are just the same as those used above.
Furthermore, $U$ is a function
only of the old coordinate $u$:
\begin{eqnarray}
U &=& - {1 \over 2} \int_u^{+\infty} {{du} \over {G^2}} \ {\rm where} \cr
G &=& e^{{1 \over 2} \int_0^u \sigma(u)du}.
\end{eqnarray}
Let us parameterize the worldline by the coordinate $U$.  Because
the connection coefficients are bounded, the ratio $dU/d\lambda$
is $C^0$ at $U=0$ for any affine parameter $\lambda$ along the worldline.
Suppose that our geodesic intersects the horizon at the point
$(U,V,q,\theta)$.  Since the curvature is continuous in $q,V,\theta^i$,
when our geodesic is close to the horizon
we may approximate the curvature by setting these coordinates equal to
their value where the worldline crosses the horizon.
Thus, $v/l \sim \int_U^0 R \ dU$.

Whether this approximation remains valid at the finite distance $U_A$
from the horizon is a more subtle question due to the multiplicity
of length scales $r_0, \sqrt{p},L$ which appear in the classical metric.
Recall, however, that due to the periodic
identifications, there are many values of $q$ corresponding to
any given point on the horizon; going once around the
$z$ direction rescales $q$ by $e^{-2\overline{\sigma}L}$.
If we use a coordinate patch in which $q$ takes values less than
$p/r_0$, then appendix A shows that the part of a generic geodesic
between $r=r_A$ and the horizon can be approximated by
taking $U$ to be an affine parameter and $U,V,q, \theta^i$ to be constant.
The qualification `generic' arises because this is in fact not
true of those geodesics which are very close to the integral curves
of the killing field ${{\partial} \over {\partial V}}$.   Such integral
curves never cross the horizon (they `run directly away from the horizon
at the speed of light'), and nearby geodesics can circle the compact
direction many times before finally crossing the horizon (so that, for
example, $q$ is not constant along those trajectories).  Such geodesics
must, however, be very carefully chosen:  given any $C^0$ timelike
vector field there is a notion of what fraction of the geodesics
fired from a given point fail to have the desired property and this fraction
vanishes for large $Q_1Q_5$.
Considering such finely tuned geodesics can thus be thought of as
a higher order correction.   The class of generic geodesics includes
{\it all} geodesics which fall toward the black string form an initial
position $r$ of order $r_A$ or larger.

As a result, the typical relative velocity imparted to test
particles with a small separation $l$ during their fall from
$U=U_A$ to some $U=U_B$ near the horizon is given by
\begin{eqnarray}
v_{wave}
/l \sim \int_{U_A}^{U_B} {{\dot{\sigma} r_0^2 } \over {r^2}}
dU \cr
 = - {{4q} \over {r_0}} \int_{u_A}^{u_B} \dot \sigma du \cr
= - {{4q} \over {r_0}} [\sigma(u_B) - \sigma(u_B)]
\end{eqnarray}
where $u_A,u_B$ are the corresponding values of the coordinate $u$.
While this does not have a well defined limit as $u_B \rightarrow \infty$,
it is clearly bounded by
\begin{equation}
|{v_{wave} \over l}| < (const) {{q} \over {r^3_0}} {{|\delta p|} \over
\sqrt{p}}
\end{equation}
where $|\delta p|$ is the scale of variation of $p(u)$.  Taking
${{|\delta p|^2} \over {p^2}} \sim  {{|\delta p_{rms}|^2} \over {p^2}}$,
we may compare this bound with the relative velocity produced
by the stationary term:
\begin{equation}
v_{stat}/l \sim \sigma^2 |U_A| = {{2q r_A^2\sqrt{p}} \over {r_0^5}}.
\end{equation}
Thus, the ratio of these effects is just
\begin{equation}
{{v_{wave}} \over {v_{stat}}} \sim {{\delta p_{rms} r_0^2} \over
{r_A^2 p}} \sim 1
\end{equation}
and,
even in the small region where the wave-induced curvature
is larger than the stationary term, it produces only about the same total
effect on the distribution of test particles.
If the particles begin their fall from
rest relative to each other at $r=r_A$,
the ratio
of the distortions is also comparable.  This relative velocity
and distortion are,
however, only small fractions of the total effect produced by
the stationary term if the test particles are dropped from $r \gg r_A$.
In this sense then, the effect of the singular term is
for a typical D-brane state.

\section{Discussion}
\label{concl}

We have seen that the ensemble is characterized
by small fluctuations (order of ${1 \over {(Q_1Q_5)^{1/2}}}$
or smaller)
in the charge distributions $p(u)$, $p_i(u)$, and $\gamma(u)$
and that the charge distributions of a typical string are very close
to that of the stationary black string.
As a result, the D-brane ensembles provide only a small statistical
correction to the horizon area and curvature of a stationary black
string, and a typical quantum state corresponds well to the classical
solution.  A similar analysis could be performed
for ensembles \cite{BLMPSV} corresponding to rotating black holes
or \cite{HM1,HM2} for a black string with nontrivial distributions
of charge.  The analysis of the horizon area should be essentially
the same in such cases, through the curvature analysis would differ
for the nonuniform case
as the classical black string has a singular horizon whenever
the charge distributions are nonuniform.

A few comments about the nature of these singular horizons
are now in order.  As noted in \cite{HY}, the singularities are
`weak' in the sense that they are twice
integrable and so produce finite distortion of any object falling through
the horizon.  In addition, we note that the connection coefficients
are bounded (see appendix)
so that geodesics crossing the horizon have well defined
tangent vectors at $U=0$ and can be uniquely continued beyond this
surface.  Note that this is true even if the wave $\sigma$ for
$U<0$ (outside) is quite different from the wave $\sigma$ for
$U>0$ (inside).  Such singularities are also
{\it dynamically} benign
as these spacetimes represent exact solutions of classical
string theory to all orders in $\alpha'$ \cite{HT} and the dilaton remains
small.  Thus, considering such singular
solutions may be physically reasonable.

The reader may note that we have not yet
discussed deviations of the
cross sections for absorption  of low energy quanta.  This is because
no new calculations are needed for this purpose.    The cross
section for absorption of a low energy quantum of energy $\omega$ has
been shown \cite{DMW,DM1,DM2} to be given (in the appropriate regime) by
\begin{equation}
{{A_H} \over {T_H}} \omega (\langle N_{\omega} \rangle + 1)
\end{equation}
where $A_H$ and $T_H$ are the horizon area and Hawking temperature
of the original stationary black hole,
and $\langle N_{\omega} \rangle$ is expectation value of the
occupation number of the mode with frequency $\omega$ in
the particular D-brane
state.  For a low energy quantum, a typical state has an occupation
number $ T_H/\omega (1 + O(1))$ as determined by the equipartition theorem.
While the fluctuations of $N_{\omega}$ for a single mode are of the same
order as the expectation value, if we consider an incoming particle with
a frequency spread much greater than the gap
associated with the black hole ($\Delta \omega \gg {{\hbar c} \over
{\tilde L}} = \omega_{gap}$) it is appropriate to average over
many modes.  A typical D-brane state will then be associated with a
cross section
\begin{equation}
A_H (1 + O( \sqrt{{\Delta \omega} \over {\omega_{gap}}})).
\end{equation}
for $\omega /T_H \ll 1$.  Here, the $O(\sqrt{{{\Delta \omega} \over
{\omega_{gap}}}})$ term represents the fluctuations.

One difficult remaining question involves the status of the singular
solutions associated with higher multipole moments ($l \ge 1$
for the waves considered here, $l \ge 2$ for external waves).  
Classically, one may argue against such solutions due to their strong
singularities.  However, a conclusive treatment has yet to be given.
One may hope to argue that the higher multipole moments of
such states vanish exactly.   Some evidence for this will be presented
in \cite{SSDM}.

Finally, some comments are in order concerning the possibility
of extending this work to non-BPS states.  Such a project would seem to
be quite difficult, not only due to the decreased protection from
supersymmetry, but also because of the fact that non-BPS classical solutions
in general lack the null translational symmetry which allowed
us to assign conserved charge {\it distributions} to each
solution and to in fact characterize each black string by its
associated distribution of charge.  Thus, the consideration of non-BPS
states destroys the bridge used here to connect the
classical and quantum descriptions.

\acknowledgements
It is a pleasure to thank Gary Horowitz, Juan Maldacena,
Samir Mathur, Cristina Marchetti,
and Rob Myers for many helpful discussions.  Special
thanks are due to Haisong Yang for supplying the unpublished
stationary curvature near the horizon.  I would also like to thank
Sumati Surya for her work in deriving related results, and Simon Ross
for pointing out an error in the Appendix in an earlier version of this
work.

\appendix

\section{Geodesics near the horizon}
\label{geo}

In section \ref{sings},
it was stated that along `generic' geodesics passing between
$U_A$ and $U=0$ the coordinates $q,V,\theta$ are approximately constant
and $U$ is effectively an affine parameter.  The purpose of this appendix
is to derive this result.  Let us begin by recalling the details of the metric
and coordinate system.  Here, we continue to specialize to the case
$r_1=r_5=r_0$, $p_i(u)=0$, $\gamma(u) = 0$,
but we restore the dimensional factors that were
suppressed in \cite{HY}.  Below, we have adjusted the normalizations
so that all three coordinates $(U,V,q)$ have dimensions of length.  Note that
our coordinate $V$ coincides with the $V$ coordinate of \cite{HY}, which
was called $\nu$ in \cite{HM1}.  Also note that our coordinate $r$ is
the $r$ coordinate of \cite{HM1} (which vanishes on the horizon)
and not the $r$ coordinate of \cite{HY}.

The coordinates $U,V,q$ are defined by
\begin{eqnarray}
\label{coords}
G(u) &=& e^{\int_0^u \sigma du} \cr
U &=& - \int_u^{+\infty} G^{-2} du \cr
W &=&  {{Gr} \over {r_0\sqrt{r^2 + r_0^2}}} \cr
q &=& -{1 \over {2r_0W^2}} - 3 r_0 \int_0^U \sigma dU \cr
V &=& v - {{\sigma r_0^4} \over {r^2}} - 2r_0^2 \int_0^u \sigma^2 du + 3
r_0^4 \int_0^U \sigma^2 W^2 dU
\end{eqnarray}
where the integrals $dU$ are performed along contours of constant $q,V,\theta$.
The metric is then
\begin{eqnarray}
ds^2 &=& - r_0^2 W^2 dUdV + r_0^4 \sigma^2 W^4 r^2 \left ( 1 + 8 r^2/r_0^2
+ 4 r^4/r_0^4 \right) dU^2 \cr
&+& \left[ 2 {{\sigma} \over {r_0}} W^4 r^2 (r^2 + r_0^2) (2 r^2 + 3 r_0^2)
+ 6W^2 r_0^7 \int_0^U \sigma^2 W^4 dU \right] dUdq \cr
&+& r_0^{-2} W^4 (r^2 + r_0^2)^3 dq^2 + (r^2 + r_0^2) d\Omega^2_3.
\end{eqnarray}

Now, for the case of interest, we have shown that the momentum density
is `slowly varying' so that the approximation $\sigma = \sqrt{p(u)}/r_0^2$
is valid.  Furthermore, we have seen that the variations $\delta p$ in
the momentum density $p(u)$ are small compared to the average value $p$.
Under these circumstances,
the form of the metric greatly simplifies near
the horizon.  Let us suppose that $\epsilon = r/r_0 \ll 1$.  Then the
corresponding value of $U$ is $U = - {{ \epsilon^2}  \over {W^2 \sqrt{p}}}
(1 + O(\epsilon^2))$.  As a result, to leading order we have $W^2 = -{1 \over
{2qr_0}}$ so that $W$
is constant along the contours of integration in (\ref{coords})
and $U_A = {{qr_0 \epsilon^2} \over {\sqrt{p}}}$.
Expanding the various terms, we may write the
metric as
\begin{equation}
ds^2  =  -r_0^2 W^2 dU dV +  r_0^6 W^4 \sigma^2  \epsilon^2 dU^2 +
\epsilon^2 \sigma W^4 r_0^4 dq dU +
W^4 r_0^4 dq^2 + r_0^2 d\Omega^2_3 + O(\epsilon^4) + O({{\delta p} \over p}),
\end{equation}
which is a much more convenient expression to work with.  It is important
to keep track of the $\epsilon^2$ terms as, if we take a derivative with
respect to $U$, ${{\partial \epsilon^2} \over {\partial U}}$ is of order
$1$.

If we now consider a null or timelike geodesic which runs from $U_A$ to
the horizon, we can use causality to bound the change in $q$ along
this path.
If we parameterize the geodesic by the coordinate $U$,
the tangent vector $n^\alpha = dx^\alpha/dU$ must satisfy
\begin{equation}
n^U n^V >  W^2 r_0^2 (n_q)^2
\end{equation}
plus corrections of order $\epsilon^2$.
Let us introduce
the parameter $n^U/n^V = \alpha^2$.  Then
we may bound $n^q$ in terms of $n^U$.  This means that the change
$\delta q$ in the coordinate $q$ along the geodesic must satisfy
\begin{equation}
\delta q < {{U_A} \over  {\alpha W r_0}} \sim {{
q^{3/2} r_0^{1/2} \epsilon^2}
\over {\alpha \sqrt{p}}}.
\end{equation}
Choosing $q < p/r_0$ yields $\delta q \sim q \epsilon^2/\alpha$, so that
unless $\alpha \sim \epsilon^2$ the coordinate $q$ changes
by only a very small amount along the geodesic.

Let us allow $\delta q \sim q$, or $\alpha \sim \epsilon^2$.
We would like to say that the excluded geodesics are but a small fraction of
the whole.  An obvious difficulty is the lack of a Lorenz
covariant normalized measure on the space of timelike and null
vectors.  Let us, however, suppose that we fix an arbitrary normalized
measure on $[1,0)$.  Then, given a normalized future directed
timelike vector $t^\alpha$ at $x$, we may use this vector to
define a measure on the space of future directed timelike geodesics
through $x$, say, by pull-back through $-t \cdot v$ where
$v$ is the unit tangent to the geodesic.  Then, one statement
that can be made is that any normalized timelike $C^0$ vector
field on the stationary black string spacetime must differ from
$|{{\partial} \over {\partial U}} +
{{\partial} \over {\partial V}}|^{-1} ( {{\partial} \over {\partial U}}
+ {{\partial} \over {\partial V}})$ by some fixed finite boost on
the horizon.  As a result, if we take the limit $Q_1,Q_5 \rightarrow
\infty$ while holding the vector field fixed, the measure assigned
to the set of excluded geodesics vanishes in that limit, no matter
what vector field was chosen.  In this
sense, then, consideration of the excluded geodesics is equivalent
to a higher order correction.  It is comforting to note that, in
terms of the original coordinates $(u,v,r)$ and any parameter
$\lambda$,  we have
\begin{equation}
{{dU} \over {d\lambda}} = G^{-2} {{du} \over {d\lambda}}
\end{equation}
\begin{equation}
{{dq} \over {d \lambda}} = -  {q \over {r_0}} {d \over {d \lambda}}
\ln(r/r_0) - q \sigma {{du} \over {d\lambda}} + O(\epsilon^2)
\end{equation}
so that any geodesic which begins at $r\ge r_A$ with $dr/d\lambda \le 0$
has ${{dq} \over {dU}} \sim {1 \over {\epsilon^2Wr_0}}$ or smaller
(corresponding to $\alpha \sim \epsilon^2$ or larger) and has been included
in our analysis.

The last statement to be verified is that $U$ may be treated as an
affine parameter along the geodesic all the way from the horizon to
$U=U_A$.   To do so, let us first note that (since
$q < p/r_0$) all of the connection coefficients $\Gamma^U_{\beta \gamma}$
are of order $\epsilon^2$ except for $\Gamma^U_{UU} \sim \sqrt{p}/r_0q$
and $\Gamma^U_{qU} \sim r_0W^2$.
As a result, since $\dot{q} \sim \epsilon^2$,
the geodesic equation says that if $\dot{U}$ denotes
the derivative of $U$ along the worldline with respect to some affine
parameter, then $d\dot{U}/\dot{U} \sim \sqrt{p}/r_0 q
dU$.  Thus the total fractional change in $\dot{U}$
is on the order of $ \sqrt{p}/r_0q U_A = 2 \epsilon^2$ and we may
treat $U$ as an affine parameter for $ 0 > U > U_A$.

\end{document}